%% file: main.tex
\documentclass[sigconf]{acmart}

\AtBeginDocument{%
  }

\usepackage{multirow}
\usepackage[ruled,linesnumbered]{algorithm2e}
\usepackage{amsmath}
\usepackage{subcaption}
\usepackage{color,xcolor,colortbl}
\usepackage{enumitem}
\usepackage{graphicx}
\usepackage{tcolorbox}
\usepackage{bbding}
\usepackage{microtype}
\usepackage{tikz}   
\usepackage{setspace}
\usepackage{bm}
\tcbuselibrary{skins} 
\usepackage{wrapfig}

\begin{document}

\newtcolorbox{myquestionbox}[1][]{
    width=.48\textwidth,
    colback=black!5,
    colframe=black!75,
    boxrule=0.5pt,
    arc=2mm,
    fonttitle=\small\bfseries,
    coltitle=black,
    top=2mm, bottom=2mm,
    left=3mm, right=3mm,
    enhanced,
    drop shadow={black!50!white, opacity=0.5},
    #1 
}

\newtcolorbox{myfindbox}[1][]{
    width=.48\textwidth,
    colback=black!5,
    colframe=black!75,
    boxrule=0.5pt,
    arc=2mm,
    fonttitle=\small\bfseries,
    coltitle=black,
    top=2mm, bottom=2mm,
    left=3mm, right=3mm,
    enhanced,
    drop shadow={black!50!white, opacity=0.5},
    #1 
}

\definecolor{mygreen}{HTML}{E3F2D9}
\definecolor{myblue}{HTML}{D4F4F1}
\definecolor{mypink}{HTML}{FADBDF}



\title{Evaluating Repository-level Software Documentation via Question Answering and Feature-Driven Development}






\author{Xinchen Wang$^{1}$, Ruida Hu$^{1}$, Cuiyun Gao$^{1 \dagger}$, Pengfei Gao$^2$, Chao Peng$^{2 \dagger}$}

\affiliation{ 
  \institution{$^1$ Harbin Institute of Technology, Shenzhen, China}
  \country{}
  }
\affiliation{%
  \institution{$^2$ Independent Researcher, China}
  \country{}
  }

\email{200111115@stu.hit.edu.cn,  
200111107@stu.hit.edu.cn,
gaocuiyun@hit.edu.cn,
}
\email{gaopf1995@gmail.com,
chao.peng@acm.org
}
\thanks{$^{\dagger}$ Corresponding authors.}


\newcommand{\benchmark}{SWD-Bench\xspace}
\newcommand{\moduleA}{High-quality data crawling and filtering stage\xspace}
\newcommand{\moduleB}{Repository-level context retrieving stage\xspace}
\newcommand{\moduleC}{Functionality-driven QA construction stage\xspace}

\newcommand{\ie}{\textit{i}.\textit{e}.\xspace}
\newcommand{\wxc}[1]{\textcolor{blue}{{#1}}}
\newcommand{\rdhu}[1]{\textcolor{darkgreen}{{#1}}}
\newcommand{\pf}[1]{\textcolor{red}{{PF: #1}}}
\newcommand{\pc}[1]{\textcolor{cyan}{{Chao: #1}}}
\definecolor{darkgreen}{rgb}{0,0.5,0}

\begin{abstract}
\label{sec:abstract}
\input{Sections/0_abstract}

\end{abstract}



\maketitle

\section{INTRODUCTION}
\label{sec:introduction}
\input{Sections/1_Introduction}

\section{TASK FORMULATION}
\label{sec:task_formulation}
\input{Sections/9_task_formulation}

\section{METHODOLOGY}
\label{sec:architecture}
\input{Sections/3_Method}

\section{EXPERIMENTAL SETUP}
\label{sec:evaluation}
\input{Sections/4_Evaluation}

\section{EXPERIMENTAL RESULTS}
\label{sec:experimental_result}
\input{Sections/5_Experimental_Result}

\section{DISCUSSION}
\label{sec:discussion}
\input{Sections/6_Discussion}

\section{RELATED WORK}
\label{sec:related}
\input{Sections/7_Related_Work}

\section{CONCLUSION}
\label{sec:conclusion}
\input{Sections/8_Conclusion}

\bibliographystyle{ACM-Reference-Format}
\bibliography{Citation} 
\end{document}

%% file: Sections/0_abstract.tex
Software documentation, which provides detailed and clear descriptions of source code, is crucial for repository comprehension. Researchers have developed various automated methods, as manual documentation writing is labor-intensive. With the advancement of large language models (LLMs), these methods extend from isolated code snippets to the entire repository, leveraging global semantic context for comprehensive summaries. However, existing benchmarks for evaluating software documentation suffer from two fundamental limitations. (1) They lack repository-level analysis, assessing in a fragmented manner that overlooks overall documentation quality. (2) They depend on unreliable evaluation strategies. While LLM-as-a-judge methods are widely adopted, their reliability is compromised by vaguely defined evaluation criteria and limited repository-level knowledge.

To address these limitations, we propose a novel benchmark for evaluating repository-level \textbf{S}oft\textbf{W}are \textbf{D}ocumentation, named \textbf{\benchmark}. Our evaluation strategy is inspired by documentation-driven development, where higher-quality documentation enables more effective repository comprehension. Based on this strategy, we propose to regard LLMs as repository developers and evaluate the documentation quality through the process of the LLMs' understanding and implementing functionalities, instead of directly prompting LLMs for evaluation. To ensure the reliability of evaluation results, documentation quality is assessed by functionality-driven question answering (QA) tasks.
Specifically, \benchmark introduces three interconnected QA tasks: (1) \textbf{Functionality Detection}, aiming to assess whether a specific functionality exists in the documentation.
(2) \textbf{Functionality Localization}, aiming to evaluate the capability in accurately locating functionality-related files. (3) \textbf{Functionality Completion}, aiming to measure the comprehensiveness of implementation details of the functionalities.
The construction pipeline for \benchmark involves three stages: we first mine high-quality Pull Requests (PRs) through multi-step filtering, then enrich them with diverse repository-level context, and finally leverage this rich context to meticulously craft the tasks. This rigorous process yields the final benchmark of 4,170 entries across three QA tasks. Extensive experiments reveal that there still exist limitations in current repository-level documentation generation methods, and highlight that source code offers complementary value to software documentation. Besides, documentation generated by the best-performed method improves the issue-solving rate of SWE-Agent, one popular issue-fixing approach, by 20.00\%, highlighting the practical value of high-quality documentation in supporting documentation-driven development.

%% file: Sections/1_Introduction.tex
\begin{figure*}[t]
	\centering
	\includegraphics[width=.8\textwidth]{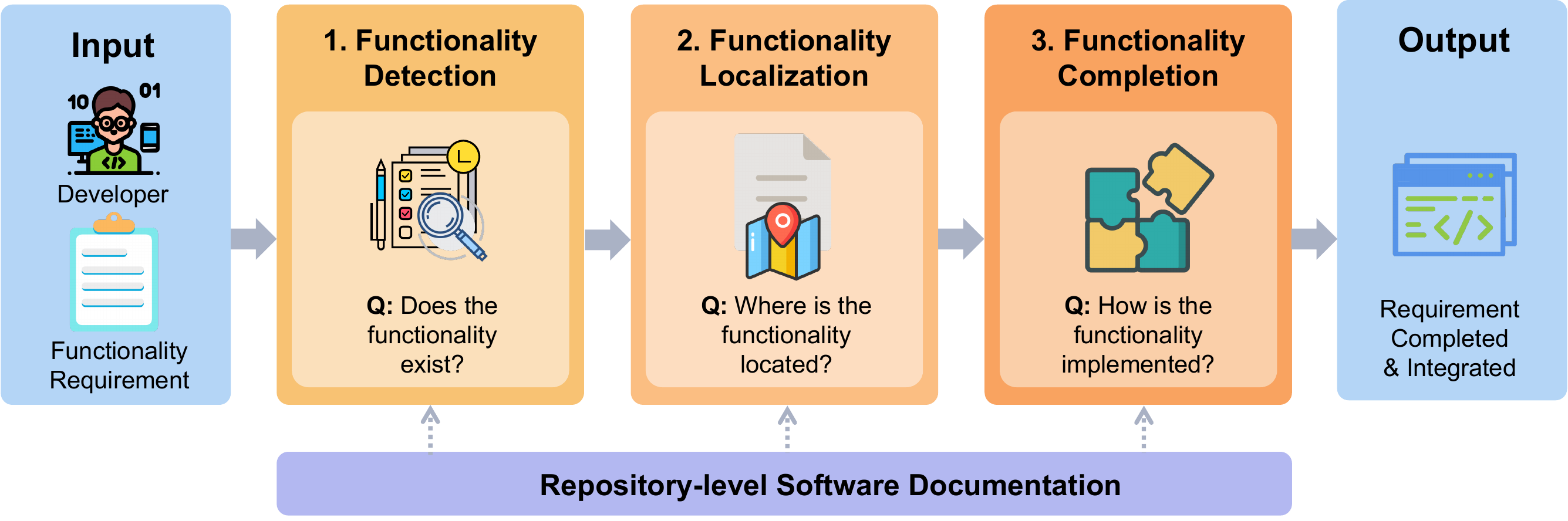}
    \vspace{-1.2em}
    \caption{A typical workflow of documentation-driven development.}
\label{fig:workflow}
\vspace{-1.5em}
\end{figure*}

Software documentation refers to describing the functionality and logic of source code, playing a crucial role in software engineering practices~\cite{survey1, survey4, survey5}. It facilitates developers' understanding of repositories, thereby enhancing development efficiency~\cite{survey2, survey3, heeager2012introducing, sommerville2001software, chomal2014significance}. Since manually writing documentation is costly, researchers have developed various automated methods. Early methods~\cite{template1, template2, info-retrieve1, info-retrieve2} are limited to summarizing isolated code snippets, such as functions or classes. They often neglect contextual information, including program dependency and functional interaction, resulting in fragmented documentation that provides limited guidance. With the advancement of large language models (LLMs), research has shifted towards generating repository-level documentation~\cite{docagent, repoagent}. These methods enhance documentation quality by retrieving semantic context from the entire repository to generate comprehensive summaries. AI-based assistants such as DeepWiki~\cite{deepwiki} and Autodoc~\cite{autodoc}, pre-generate software documentation for the user repository, thereby empowering diverse code tasks, such as code generation and issue solving.

Despite the progress in repository-level software documentation generation, existing benchmarks~\cite{repoagent, docagent, hierarchical} suffer from two fundamental limitations that hinder comprehensive evaluation: \textbf{(1) Lack of repository-level analysis:} Current benchmarks typically decompose generated documentation into function- or method-level summaries and assess them sequentially, overlooking semantic relationships between code snippets. This narrow focus does not well reflect real-world documentation reading, where developers rely on a holistic understanding to capture functionality spanning multiple modules. Hence, these benchmarks fail to assess the overall documentation accuracy. \textbf{(2) Unreliable evaluation strategy:} Current evaluation strategies generally fall into three types. Human-based methods~\cite{human1, human2} are labor-intensive, while metric-based methods~\cite{bleu,meteor,rouge} rely on reference documentation, which is generally difficult to construct. Hence, LLM-as-a-judge methods~\cite{docagent, hierarchical} are widely adopted due to their strong contextual understanding abilities. These methods typically utilize an LLM to assess documentation quality through a 5-point Likert scale~\cite{likert}. However, these scales rely on vague descriptors like ``not helpful'' and ``slightly helpful'', rather than a precise and objective definition. Besides, the LLM often lacks domain knowledge of the repository's implementation details, making it difficult to determine whether a documented functionality is accurately described. Furthermore, these strategies tend to assess the qualities of the documentation content, rather than its practical utility.

To mitigate these limitations, we propose a novel benchmark for evaluating repository-level \textbf{S}oft\textbf{W}are \textbf{D}ocumentation, named \textbf{\benchmark}. Our evaluation strategy is inspired by documentation-driven development~\cite{1377190, heeager2012introducing, survey2}, where higher-quality documentation enables developers to more effectively understand the repository. Figure~\ref{fig:workflow} illustrates a typical development workflow: when encountering functionality requirements, developers begin by searching the documentation to analyze whether the functionality has been implemented. If so, they then map the documentation descriptions to specific locations within the vast repository. Finally, developers leverage concrete information—such as API definitions and parameter usage—provided in the documentation to integrate the new functionality. Clearly, software documentation lies at the core of these consecutive stages. Based on this workflow, we simulate the LLM as a repository developer that understands and implements functionalities through a documentation-based inquiry process, instead of directly scoring the documentation. Accordingly, we construct three categories of repository-level, objective question-answering (QA) tasks aligned with this development workflow: \textbf{Functionality Detection}, \textbf{Functionality Localization}, and \textbf{Functionality Completion}. The documentation quality is measured by assessing the LLM's performance on these tasks. 

The construction pipeline of \benchmark is divided into three stages. (1) \textbf{\moduleA:} Pull Requests (PRs) typically introduce functionalities and contain rich contextual information, making them ideal for QA construction. Hence, we employ a series of rigorous crawling and filtering rules to retain high-quality PRs that genuinely reflect developers' functional contributions from representative repositories. (2) \textbf{\moduleB:} We enrich each PR with extensive contextual information, covering its background, motivation, and impact scope, such as program dependencies and associated issues. This comprehensive context provides a solid foundation for constructing repository-level QA tasks, thereby mitigating the limitation of insufficient analysis from a holistic perspective. (3) \textbf{\moduleC:} We leverage diverse context from PRs to meticulously create three categories of QA tasks. Each question is paired with a clear reference answer extracted from the PR's context. Based on this, we mitigate the limitation of unreliable evaluation strategies by assessing software documentation quality through the consistency between the LLM’s answers and the objective reference answers. \benchmark consists of 4,170 high-quality entries across three tasks. To ensure data quality, we manually validate a random sample of 100 entries, achieving a Kappa coefficient greater than 90\%, which indicates strong inter-annotator agreement. 
 
We conduct extensive experiments and conclude several findings:
\begin{enumerate}[leftmargin=*]
    \item There still exist limitations in current repository-level software documentation generation methods, among which methods leveraging more thorough context achieve better performance.
    \item Software documentation produced by the best method improves SWE-Agent's issue-solving rate by 20.00\%, highlighting its practical value in supporting documentation-driven development.
    \item Source code offers complementary value to software documentation on repository comprehension, especially in functionality detection and localization. 
\end{enumerate}

Our contributions can be summarized as follows:
\begin{enumerate}[leftmargin=*]
    \item We introduce \benchmark, a novel benchmark for evaluating repository-level software documentation. This benchmark aims to mitigate two major limitations, including the lack of repository-level analysis and unreliable evaluation strategies.
    \item \benchmark comprises 4,170 high-quality data entries. Each entry is enriched with three categories of functionality-driven QA tasks, enabling holistic and comprehensive evaluation of software documentation quality.
    \item We conduct extensive experiments on \benchmark, conclude our findings, and provide valuable insights for both researchers and developers.
\end{enumerate}

%% file: Sections/9_task_formulation.tex
\begin{figure*}[t]
	\centering
	\includegraphics[width=\textwidth]{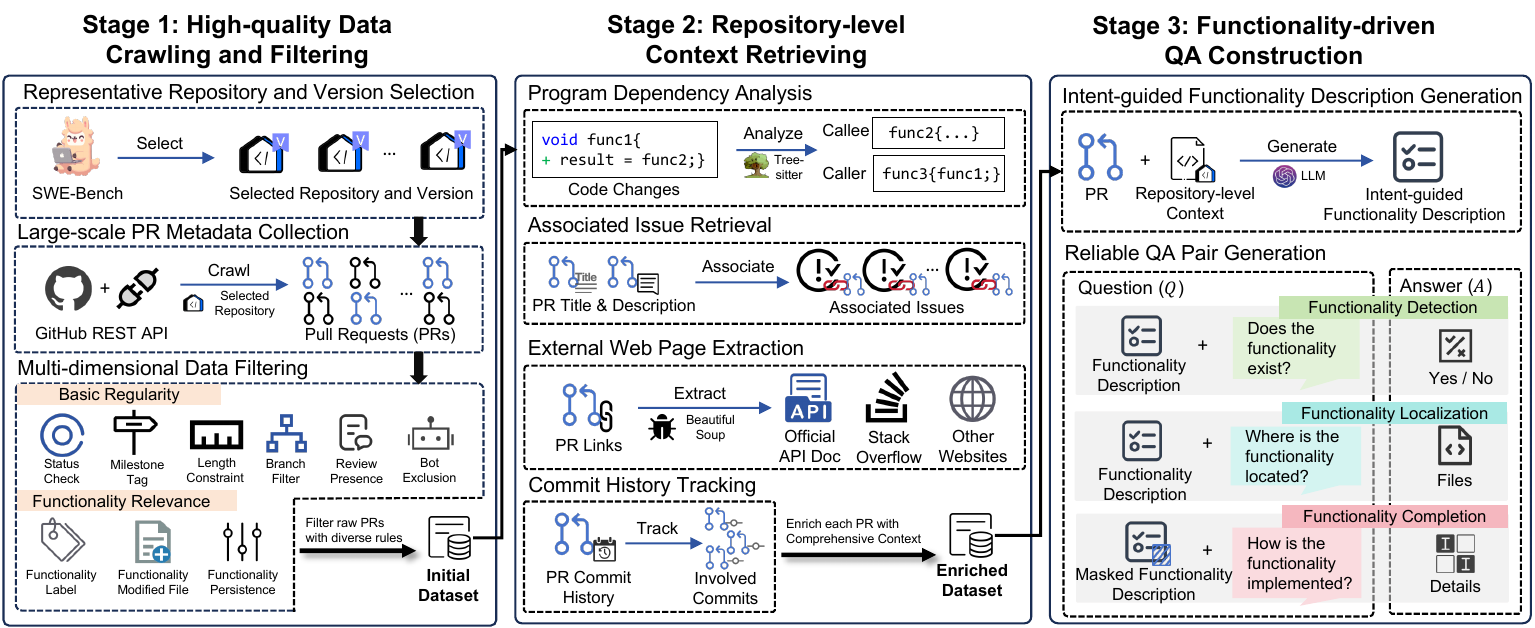}
    \vspace{-2em}
    \caption{The overview of \benchmark's data construction pipeline.}
\label{fig:framework}
\vspace{-1em}
\end{figure*}

We formulate the evaluation on \benchmark as a documentation-based QA problem. Let $\mathcal{D}$ denote the space of generated software documentation and $\mathcal{Q}$ be the set of potential developers' questions. The target model is defined as a function $M: \mathcal{D} \times \mathcal{Q} \rightarrow \mathcal{A}$, which takes the documentation $D \in \mathcal{D}$ and a specific question $Q \in \mathcal{Q}$ as input, and produces an answer $A \in \mathcal{A}$. The structure of the question $Q$ and the domain of the answer space $\mathcal{A}$ vary across the three tasks, as detailed below.

\subsection{Functionality Detection}
    Software documentation supports developers in identifying which functionalities are implemented within the repository. This task reflects documentation’s completeness in presenting repository functionalities. In this task, the question $Q_{\text{detect}}$ inquires about the existence of a specific functionality. The answer space $\mathcal{A}$ is binary, \ie, $\mathcal{A} = \{\text{True}, \text{False}\}$. The model output $y = M(D, Q_{\text{detect}})$ indicates whether the functionality is judged to be available.

\subsection{Functionality Localization}
    Software documentation helps developers effectively locate the source files responsible for implementing specific functionalities. This task reflects the documentation’s helpfulness in navigation and localization. Here, the question $Q_{\text{localize}}$ asks for the implementation location of a functionality. The answer space $\mathcal{A}$ corresponds to the power set of all file paths in the repository. The model predicts a list of files $F = M(D, Q_{\text{localize}})$ responsible for the functionality.

\subsection{Functionality Completion}
    Software documentation enables developers to obtain clear technical details about specific functionalities. This task reflects the documentation’s comprehensiveness in describing the functionalities' implementation details. Specifically, the question $Q_{\text{complete}}$ is formulated as a cloze-style prompt containing masked placeholders. The answer space $\mathcal{A}$ consists of sequences of details (e.g., API parameters). The model generates the missing details $T = M(D, Q_{\text{complete}})$ to complete the masked placeholders.

%% file: Sections/3_Method.tex
This section details the data construction pipeline for \benchmark, which is illustrated in Figure~\ref{fig:framework}.

\subsection{High-quality Data Crawling and Filtering Stage}
To ensure high-quality and large-scale data collection, we follow a multi-step data mining process.

\subsubsection{Representative Repository and Version Selection}
\label{sec:repo_selection}
To construct a robust benchmark, we follow the repository selection strategy established by SWE-Bench~\cite{swebench}, a widely recognized issue-resolving benchmark encompassing repositories from diverse application domains with active development communities. Specifically, we use 12 repositories from SWE-Bench as the sources for PR metadata collection. For each repository, we designate a single snapshot as the evaluation target: the latest stable version available in SWE-Bench (2023). Thus, our benchmark comprises \textbf{12 distinct repository versions} (one per repository) for documentation generation.

\subsubsection{Large-scale PR Metadata Collection}
We leverage the GitHub REST API~\cite{githubapi} to crawl all PRs from the selected repositories to ensure a reliable collection. For each PR, we systematically extract and structure essential metadata, such as its unique ID, title, description, and code changes. This initial collection process yields a massive corpus of 177.4k PRs, which are then stored in a structured format for subsequent filtering.

\subsubsection{Multi-dimensional Data Filtering}
Raw crawled data is often noisy. To guarantee the high quality of our benchmark, we apply a multi-step filtering process.

\textbf{\textit{Basic Regularity Filtering:}} This step aims to quickly filter noisy PRs from the initial dataset. The specific rules are as follows:
\begin{itemize}[leftmargin=*]
    \item \textbf{Status Check:} Retain PRs that have been merged, as these represent accepted contributions that meet repository standards.
    \item \textbf{Milestone Tag:} Retain PRs containing the ``milestone'' attribute, as these are typically associated with major repository goals.
    \item \textbf{Length Constraint:} Retain PRs with description length greater than 50 characters, ensuring each PR provides clear context.
    \item \textbf{Branch Filter:} Retain PRs merged into the main branch, as these are formal contributions central to the repository's evolution.
    \item \textbf{Review Presence:} Retain PRs with review comments, indicating that the changes have undergone human validation.
    \item \textbf{Bot Exclusion:} Exclude PRs generated by bots to focus on human-made changes.
    
\end{itemize}

\textbf{\textit{Functionality Relevance Filtering:}}
This step aims to retain PRs focused on functionality implementation. 

\begin{itemize}[leftmargin=*]

    \item \textbf{Functionality Label:} Retain PRs that contain functionality-related labels, such as ``new feature'' or ``new API'', to ensure the relevance to functional contributions.
    \item \textbf{Functionality Modified File:} Retain PRs based on the filenames of the modified files. Specifically, a PR is kept if it modifies at least one functional file (such as ``.py'' extension) located within a functional directory (excluding directories like ``\texttt{/test}'').
    \item \textbf{Functionality Persistence:} Functionalities introduced by previous PRs may be altered or removed in subsequent code updates. To ensure the persistence of QA tasks, we focus on PRs merged before the snapshot time of the selected repository version in Section~\ref{sec:repo_selection}. We then verify the persistence of their introduced functionalities by checking whether the added non-comment code lines still exist in the corresponding files of the selected version, accounting for potential file renames.

\end{itemize}
After this series of rigorous filtering steps, we obtain 4,170 high-quality PRs, forming the comprehensive basis of our benchmark.

\subsection{Repository-level Context Retrieving Stage}
\label{sec:stage2}

Current benchmarks often assess in a fragmented manner, failing to evaluate overall accuracy. To address this, this stage aggregates abundant repository-level context for each PR, forming a solid foundation for constructing high-quality QA tasks.

\subsubsection{Program Dependency Analysis}

Code changes in PRs often propagate their impact beyond the immediately modified snippets, potentially affecting the dependent modules. Since raw diffs are fragmentary, we first utilize Tree-Sitter~\cite{tree-sitter} to parse and extract complete definitions of modified code snippets (e.g., functions, methods). To capture repository-level context, we analyze dependencies by identifying both the callers and callees of these snippets. 

\subsubsection{Associated Issue Retrieval}
Associated issues reflect repository requirements or feature motivations driving PR changes. We apply keyword-based regular expressions (e.g., \texttt{``closes''}) to extract related issue numbers from PR titles and descriptions, and then crawl issue metadata via the GitHub REST API. For the ``Django'' repository, where issues are tracked on its official website, we implement a custom crawler to ensure comprehensive coverage. This process grounds QA tasks in the broader repository context.

\subsubsection{External Web Page Extraction}
External web links in PR descriptions provide additional information, such as official documentation or community discussions. Using the BeautifulSoup package~\cite{BeautifulSoup}, we parse these pages and extract relevant information, supporting a more comprehensive understanding of the PR.

\subsubsection{Commit History Tracking}
PR typically consists of a series of commits, each documenting incremental changes. For every involved commit, we collect detailed metadata including commit message, code changes, and associated review comments. This information offers a holistic view of functionality evolution, deepening the overall understanding of PRs.

\subsection{Functionality-driven QA Construction Stage}

Current benchmarks often use vague scoring criteria, leading to unreliable evaluation. To overcome this, we leverage the rich PR context to construct functionality-driven tasks, enabling objective evaluation by comparing the LLM's answers with references. The core philosophy is to simulate documentation-driven development: a developer formulates a functionality requirement and consults the documentation for answers. Thus, each QA consists of two parts: 

\noindent \textbf{$\blacktriangleleft$} \textbf{\textcolor{violet}{Question}}: A composite input (developer's requirement) containing a \textbf{Functionality Description} and a \textbf{Query} (e.g., ``Determine if the functionality is implemented in the current repository?'').

\noindent \textbf{$\blacktriangleright$} \textbf{\textcolor{orange}{Answer}}: The factual answer extracted from the PR's metadata and context. Crucially, the functionality description serves as the content of the question, and LLMs should use this description to reason over the given documentation and predict the answer.

\subsubsection{Intent-guided Functionality Description Generation}
To generate high-quality functionality descriptions, we leverage LLMs' advanced contextual understanding capabilities. Specifically, we populate a predefined prompt template with the PR metadata and rich context from Section~\ref{sec:stage2}. We further employ the Chain-of-Thought (CoT) strategy~\cite{chain}, guiding the LLM to consider from a global perspective, which includes motivation, implementation details, and impact scope. The generated descriptions are structured along three intent dimensions to include technical details:
\begin{itemize}[leftmargin=*]
    \item \textbf{WHAT:} Entities constituting or affected by the functionality.
    \item \textbf{WHY:} Purpose and motivation behind the functionality.
    \item \textbf{HOW:} Detailed approaches used for implementation.
\end{itemize}
Here, \textbf{WHAT} and \textbf{HOW} can be derived from code changes and program dependencies, while \textbf{WHY} can be informed by commit messages and associated issues. Other contextual information further enriches the generated descriptions. 
To prevent answer leakage, the LLM is instructed to avoid explicit mentions of file paths or repository versions. Overall, the generated description simulates the information developers seek in the documentation.

\subsubsection{Reliable QA Pair Generation}
Based on the generated functionality descriptions, we formulate three kinds of QA tasks.

\paragraph{\textbf{Functionality Detection}}
This task focuses on detecting the presence of the described functionality in the current repository.
\begin{description}[leftmargin=0pt, labelwidth=2em]
    \item[\textbf{$\blacktriangleleft$} \textbf{\textcolor{violet}{Question}}:] We formulate the following question by combining the functionality description with a query:
\begin{myfindbox}
    \small \textit{\textbf{[Functionality Description]} + Determine if the functionality is implemented in the current repository?}
\end{myfindbox}
    \item [\textbf{$\blacktriangleright$} \textbf{\textcolor{orange}{Answer}}:] We compare the PR's merged time with the snapshot time of the selected repository version. The answer is \textbf{True} if the attribute is earlier than or the same as the target version, indicating the functionality is present; otherwise, the answer is \textbf{False}.
\end{description}

\paragraph{\textbf{Functionality Localization}}
This task focuses on locating the files responsible for implementing the described functionality.
\begin{description}[leftmargin=0pt, labelwidth=2em]
    \item[\textbf{$\blacktriangleleft$} \textbf{\textcolor{violet}{Question}}:] We formulate the following question by combining the functionality description with a query:
\begin{myfindbox}
    \small \textit{\textbf{[Functionality Description]} + Identify the code file(s) responsible for implementing the functionality?}
\end{myfindbox}
    \item [\textbf{$\blacktriangleright$} \textbf{\textcolor{orange}{Answer}}:] We generate the reference answer by extracting files from the PR's code changes, retaining functional files (such as \texttt{``.py''} extension) with newly added lines that are not located in non-functional directories (such as \texttt{``/docs''} or \texttt{``/tests''}).
\end{description}

\paragraph{\textbf{Functionality Completion}}
This task focuses on filling in the masked technical details of the described functionality.
\begin{description}[leftmargin=0pt, labelwidth=2em]
    \item[\textbf{$\blacktriangleleft$} \textbf{\textcolor{violet}{Question}}:] We construct the question by replacing technical details from the \textbf{WHAT}, \textbf{WHY}, and \textbf{HOW} dimensions of the functionality description with ``\texttt{[MASK]}'', and appending the query:

\begin{myfindbox}
    \small \textit{\textbf{[Masked Functionality Description]} + Fill in the \texttt{[MASK]} placeholders with the correct details?}
\end{myfindbox}
    \item [\textbf{$\blacktriangleright$} \textbf{\textcolor{orange}{Answer}}:] The reference answer is the list of corresponding technical details that are extracted from the functionality description.
\end{description}

As shown in Table~\ref{tab:benchmark_stats}, each entry in \benchmark features a detailed functionality description with an average length of 771.45 characters. On average, solving the questions in the entry requires locating 2.01 files and completing 7.48 details\footnote{Due to space limitations, the detailed prompt template and data structure are provided in our repository.}.

\subsubsection{Manual Quality Validation}
To validate the quality of \benchmark, we conduct a human calibration process. We randomly sample 100 entries and have them reviewed by two expert annotators, each possessing over three years of Python expertise. For the first task, annotators assess whether the existence of the described functionality aligns with the answer by carefully analyzing the code repository. For the second task, they verify that the answer (a file list) accurately corresponds to the described functionality through inspection of the code repository. For the third task, annotators ensure that the details required to fill are precise and can be sourced from PR's metadata and context. Besides, annotators evaluate whether the functionality description is clear and reasonable. Across all aspects, the inter-annotator agreement consistently exceeds a Kappa coefficient of 90\%, demonstrating high agreement.

\input{Tables/Benchmark}

%% file: Tables/Benchmark.tex
\begin{table}[t]
\centering
\small
\caption{Statistics of the \benchmark. \textbf{\# Func. Desc.} denotes the average character length of functionality descriptions. \textbf{\% Detect. Ratio} is the percentage of positive entries for the detection task. \textbf{\# Loc. File} and \textbf{\# Comp. Detail} represent the number of files to locate and details to complete, respectively.}
\vspace{-1em}
\label{tab:benchmark_stats}
\renewcommand{\arraystretch}{0.6}
\setlength{\tabcolsep}{1.2pt}
\begin{tabular}{cc c ccc ccc}
\toprule
\multirow{2}{*}{\textbf{\# Entry}} & \multirow{2}{*}{\textbf{\# Func. Desc.}} & \multirow{2}{*}{\textbf{\% Detect. Ratio}} & \multicolumn{3}{c}{\textbf{\# Loc. File}} & \multicolumn{3}{c}{\textbf{\# Comp. Detail}}\\
\cmidrule(lr){4-6} \cmidrule(lr){7-9}
& & & Min & Max & Avg. & Min & Max & Avg. \\
\midrule
4,170 & 771.45 & 50.65 & 1 & 62 & 2.01 & 3 & 23 & 7.48 \\
\bottomrule
\end{tabular}
\vspace{-1.0em}
\end{table}

%% file: Sections/4_Evaluation.tex
We investigate the following three research questions (RQs):

\begin{itemize}[leftmargin=*]
    \item RQ1: How do different software documentation generation methods perform on functionality-driven QA tasks? 
    \item RQ2: How effective is our evaluation strategy in assessing software documentation quality? 
    \item RQ3: What is the impact of software documentation quality on issue solving?
\end{itemize}

\subsection{Selected Methods}
To comprehensively evaluate different software documentation, we compare six approaches. Among them, four are widely adopted or academically recognized repository-level software documentation generation methods, while the remaining two serve as baselines.

\subsubsection{Baseline Methods}
\begin{itemize}[leftmargin=*]
    \item \textbf{Human-Written Documentation Artifacts (H-Written)} consist of documentation embedded directly within the source code, such as function docstrings and inline comments, as well as standalone documents, like ``README.md'' and ``.rst'' files.
    \item \textbf{Chat}~\cite{docagent} is a baseline approach that generates documentation for each code snippet (e.g., class and function), without providing any repository-level context to the LLM.
\end{itemize}

\subsubsection{Repository-level Documentation Generation Methods}
\begin{itemize}[leftmargin=*]
    \item \textbf{DeepWiki}~\cite{deepwiki} produces high-level, modular documentation by summarizing each module with its technology stack and interaction diagrams. It also includes the core file paths and source code responsible for the modules.
    \item \textbf{AutoDoc}~\cite{autodoc} performs a depth-first traversal to index the entire code repository. The documentation is generated for each file and folder, which can be combined to describe system components and how components work together.
    \item \textbf{DocAgent}~\cite{docagent} is a multi-agent system designed for iterative documentation generation. It orchestrates a team of specialized agents to determine a dependency-aware processing order and gather context from both internal and external sources. The documentation is generated for each code snippet, with agents iteratively writing and validating the content.
   \item \textbf{RepoAgent}~\cite{repoagent} is a three-stage method designed for context-aware documentation. The approach first conducts a global analysis to build the dependency graph (DAG), capturing the entire repository's structure. It then leverages this contextual information to prompt an LLM to generate fine-grained, structured documentation for each code snippet.
\end{itemize}

\subsection{Evaluation Metrics}

\subsubsection{Functionality Detection}
We use the following two metrics, considering that both the positive and negative classes are important in this binary task.

\textbf{Balanced Accuracy (B-ACC)} is the average of recall obtained on each class, providing a more representative measure than standard accuracy. It is calculated as:
\begin{equation}
    \text{B-ACC} = \frac{1}{2} \left( \frac{TP}{TP+FN} + \frac{TN}{TN+FP} \right)
\end{equation}
    
\textbf{Matthews Correlation Coefficient (MCC)} is a reliable metric of binary classification, particularly useful in imbalanced datasets, calculated as:
\begin{equation}
    \text{MCC} = \frac{TP \times TN - FP \times FN}{\sqrt{(TP+FP)(TP+FN)(TN+FP)(TN+FN)}}
\end{equation}

\subsubsection{Functionality Localization}
This task requires the LLM to predict a list of implementation files. We leverage the following two metrics to measure performance:

\textbf{F1 Score (F1)} provides a balanced assessment of precision and recall. We report the unweighted macro-average across all entries:
\begin{equation}
    \text{F1}_i = 2 \times \frac{|P_i \cap R_i|}{|P_i| + |R_i|}, \quad
        \text{F1} = \frac{1}{N} \sum_{i=1}^{N} \text{F1}_i
\end{equation}

\textbf{Intersection over Union (IoU)} measures the average overlap between the predicted and reference sets. This metric is calculated as follows, where $P_i$ and $R_i$ are the predicted and reference set of the $i$-th entry, and $N$ is the total number of entries:
\begin{equation}
    \text{IoU}_i = \frac{|P_i \cap R_i|}{|P_i \cup R_i|}, \quad
    \text{IoU} = \frac{1}{N} \sum_{i=1}^{N} \text{IoU}_i
\end{equation}

\subsubsection{Functionality Completion}
This task requires the LLM to predict a list of technical details. We use a thresholded Exact Match (EM) score to measure performance.

\textbf{Exact Match (EM):} 
EM computes the average proportion of correctly predicted details. For the $i$-th entry, we compare every predicted detail $P_{i,j}$ against its reference detail $R_{i,j}$. A match is counted if their edit similarity meets a specified threshold $\tau$. We use two thresholds to evaluate the predictions: $\tau=1.0$ for a strict, perfect match, and $\tau=0.8$ for a relaxed match. It is calculated as:
\begin{equation}
    \text{EM}_\tau = \frac{1}{N} \sum_{i=1}^{N} \frac{\sum_{j=1}^{|R_i|} \mathbb{I}(\text{sim}(P_{i,j}, R_{i,j}) \ge \tau)}{|R_i|}
\end{equation}

\subsection{Implementation Details}
\label{sec:impl_detail}
\textbf{Dataset Construction and Evaluation.}
We use Claude-Sonnet-4~\cite{claude} to generate intent-oriented functionality descriptions. Due to financial costs, we randomly sample a subset of 480 entries from \benchmark for evaluation. During evaluation, GPT-4.1~\cite{gpt} and Gemini-2.5-Pro~\cite{gemini} are employed as repository developers and address QA tasks. The sampling temperature for LLMs is set to 0.2. All experiments are repeated three times, and average results are reported to ensure reliability.

\textbf{Method Configuration.}
All automated methods are provided access to Claude-4-Sonnet~\cite{claude}, except for DeepWiki, which does not support model selection. These methods are implemented using their official replication packages or online platforms and executed with default hyperparameters.

\input{Tables/RQ1}

\textbf{Software Documentation Retrieval.}
In the documentation-driven development process, developers consult documentation for relevant information. To mirror real-world workflows, we adopt a retrieval-based strategy, supplying the LLM with relevant documentation context for each task question:

\begin{enumerate}[leftmargin=*]
\item \textbf{Chunking:} Documentation is divided into chunks of up to 512 tokens~\cite{sfr, coir}, with a 10\% overlap to preserve boundary context. Chunking follows the documentation structure: for approaches that generate summaries for each code snippet, chunks are based on syntax elements like classes and methods; for approaches that generate file or module-level summaries, chunks follow logical sections such as Markdown headings (``\#'', ``\#\#''). For DeepWiki, references to specific code fragments (e.g., ``\texttt{main.py 1-100}'') are replaced with the actual code.

\item \textbf{Embedding:} All chunks and task questions are encoded as vectors using the advanced SFR-Embedding-Code-400M\_R model~\cite{sfr}, ensuring precise retrieval of relevant documentation.

\item \textbf{Retrieval:} For each task question, the Top-K most relevant chunks are retrieved based on vector similarity and combined into context windows of different sizes (Top-1024, Top-2048, and Top-4096 tokens). We leverage these three sizes to simulate different levels of developer engagement with documentation. Token counting uses official packages~\cite{tiktoken, googlecloud}, and each retrieved chunk is annotated with its documentation file path.
\end{enumerate}

%% file: Tables/RQ1.tex
\begin{table*}[t]

\centering
\caption{Experimental results for the functionality-driven tasks. \textbf{Top XX} indicates the token size of the retrieved documentation context. \textbf{No Doc} refers to the setting of not inquiring about documentation. The largest and second-largest values in each column are highlighted with an \underline{underline}, and the largest value is also \textbf{bolded}.
}
\vspace{-1em}
\label{tab:rq1_results}
\setlength{\tabcolsep}{6.5pt}
\renewcommand{\arraystretch}{0.75}
\begin{tabular}{l|cc|cc|cc|cc|cc|cc}
\toprule
\multirow{2}{*}{\textbf{Model}} & \multicolumn{6}{c|}{\textbf{GPT-4.1}} & \multicolumn{6}{c}{\textbf{Gemini-2.5-pro}} \\
& \multicolumn{2}{c}{\textbf{Top 1024}} & \multicolumn{2}{c}{\textbf{Top 2048}} & \multicolumn{2}{c|}{\textbf{Top 4096}} & \multicolumn{2}{c}{\textbf{Top 1024}} & \multicolumn{2}{c}{\textbf{Top 2048}} & \multicolumn{2}{c}{\textbf{Top 4096}} \\ \midrule
\rowcolor{mygreen} \multicolumn{13}{c}{\textbf{Functionality Detection}} \\ \midrule
\rowcolor{gray!20} \textbf{Metric (\%)} & B-ACC & MCC & B-ACC & MCC & B-ACC & MCC & B-ACC & MCC & B-ACC & MCC & B-ACC & MCC \\
\midrule
No Doc            & 46.72 & -8.03 & -- & -- & -- & -- & 50.63 & 1.18 & -- & -- & -- & -- \\ \midrule
 \multicolumn{13}{c}{\textbf{Baseline Methods}} \\ \midrule
H-Written        & 53.59 & 11.74  & 54.06 & 13.70 & 55.16 & 15.38 & 58.13 & 15.71 & \underline{61.56} & \underline{22.07} & 62.97 & 24.66 \\
Chat       & 53.13 & 8.61 & 53.75 & 9.74 & 54.69 & 11.63 & 57.34 & 14.04 & 59.84 & 18.60 & 60.94 & 20.63 \\ \midrule

 \multicolumn{13}{c}{\textbf{Repository-level Documentation Generation Methods}} \\ \midrule
DeepWiki   & 52.19 & 6.94  & 52.50 & 8.01 & 53.91 & 10.26 & 53.91 & 8.48  & 55.78 & 11.80 & 56.09 & 12.15 \\
AutoDoc    & 52.97 & 10.13  & 53.28 & 12.23 & 54.22 & 12.59 & 57.19 & 13.70 & 59.06 & 17.24 & 60.78 & 20.33 \\
DocAgent   & \underline{54.69} & \underline{13.36} & \underline{54.84} & \underline{14.01} & \underline{55.63} & \underline{15.40} & \underline{59.69} & \underline{18.95} & 61.41 & 21.88 & \underline{63.13} & \underline{24.91} \\
    RepoAgent  & \underline{\textbf{62.19}} & \underline{\textbf{25.92}} & \underline{\textbf{62.97}} & \underline{\textbf{26.74}} & \underline{\textbf{63.75}} & \underline{\textbf{27.77}} & \underline{\textbf{63.28}} & \underline{\textbf{25.04}} & \underline{\textbf{67.66}} & \underline{\textbf{33.38}} & \underline{\textbf{69.53}} & \underline{\textbf{37.23}} \\

\midrule
\rowcolor{myblue} \multicolumn{13}{c}{\textbf{Functionality Localization}} \\ \midrule
\rowcolor{gray!20} \textbf{Metric (\%)} & F1 & IoU & F1 & IoU & F1 & IoU & F1 & IoU & F1 & IoU & F1 & IoU \\
\midrule
No Doc & 29.12 & 27.24 & -- & -- & -- & -- & 31.55 & 29.74 & -- & -- & -- & -- \\ \midrule
 \multicolumn{13}{c}{\textbf{Baseline Methods}} \\ \midrule
H-Written & 60.62 & 57.77 & 62.89 & 59.48 & 65.84 & 62.31 & \underline{62.02} & \underline{59.38} & \underline{65.97} & \underline{62.84} & \underline{68.17} & \underline{65.09} \\
Chat & 59.71 & 56.37 & 62.93 & 59.33 & 64.43 & 60.59 & 60.93 & 58.19 & 61.97 & 58.86 & 63.21 & 59.47 \\
\midrule
 \multicolumn{13}{c}{\textbf{Repository-level Documentation Generation Methods}} \\ \midrule
DeepWiki & 40.65 & 38.17 & 42.66 & 39.94 & 43.76 & 41.27 & 49.84 & 47.30 & 52.57 & 50.30 & 57.25 & 53.64 \\
AutoDoc & 56.87 & 53.94 & 59.64 & 56.43 & 62.58 & 59.13 & 59.55 & 56.42 & 60.54 & 57.27 & 64.72 & 61.02 \\
DocAgent & \underline{62.42} & \underline{58.91} & \underline{65.78} & \underline{62.08} & \underline{68.32} & \underline{64.89} & 61.74 & 58.75 & 63.69 & 60.20 & 65.96 & 62.15 \\
RepoAgent & \underline{\textbf{63.64}} & \underline{\textbf{60.21}} & \underline{\textbf{67.43}} & \underline{\textbf{64.04}} & \underline{\textbf{70.19}} & \underline{\textbf{66.28}} & \underline{\textbf{67.48}} & \underline{\textbf{64.61}} & \underline{\textbf{69.83}} & \underline{\textbf{66.71}} & \underline{\textbf{71.11}} & \underline{\textbf{67.52}} \\ \midrule

\rowcolor{mypink} \multicolumn{13}{c}{\textbf{Functionality Completion}} \\ \midrule
\rowcolor{gray!20} \textbf{Metric (\%)} & $\text{EM}_{1.0}$ & $\text{EM}_{0.8}$ & $\text{EM}_{1.0}$ & $\text{EM}_{0.8}$ & $\text{EM}_{1.0}$ & $\text{EM}_{0.8}$ & $\text{EM}_{1.0}$ & $\text{EM}_{0.8}$ & $\text{EM}_{1.0}$ & $\text{EM}_{0.8}$ & $\text{EM}_{1.0}$ & $\text{EM}_{0.8}$ \\
\midrule
No Doc      & 17.14 & 19.06 & -- & -- & -- & -- & 19.66 & 22.41 & -- & -- & -- & -- \\ \midrule
 \multicolumn{13}{c}{\textbf{Baseline Methods}} \\ \midrule
H-Written  & 24.54 & 27.23 & 25.00 & \underline{27.81} & \underline{25.58} & \underline{28.25} & 26.61 & 30.62 & 27.63 & \underline{31.39} & 28.09 & \underline{31.54} \\
Chat     & 23.43 & 26.28 & 24.25 & 26.74 & 25.38 & 27.82 & 25.36 & 27.79 & 27.38 & 30.54 & 27.85 & 30.70 \\
\midrule
 \multicolumn{13}{c}{\textbf{Repository-level Documentation Generation Methods}} \\ \midrule
DeepWiki & 22.23 & 24.45 & 22.78 & 24.74 & 23.16 & 25.53 & 25.44 & 28.47 & 26.23 & 29.85 & 26.37 & 29.73 \\
AutoDoc  & 22.88 & 25.39 & 23.41 & 26.33 & 23.77 & 27.02 & 26.00 & 29.54 & 26.23 & 29.57 & 27.09 & 30.35 \\
DocAgent & \underline{24.92} & \underline{27.61} & \underline{25.04} & 27.59 & 25.42 & 28.05 & \underline{28.09} & \underline{31.29} & \underline{28.30} & 30.82 & \underline{28.26} & 30.56 \\
RepoAgent & \underline{\textbf{26.40}} & \underline{\textbf{29.37}} & \underline{\textbf{26.06}} & \underline{\textbf{28.53}} & \underline{\textbf{26.87}} & \underline{\textbf{29.72}} & \underline{\textbf{29.40}} & \underline{\textbf{33.20}} & \underline{\textbf{29.76}} & \underline{\textbf{32.95}} & \underline{\textbf{30.05}} & \underline{\textbf{33.88}} \\

\bottomrule
\end{tabular}
\vspace{-1em}
\end{table*}

%% file: Sections/5_Experimental_Result.tex
\subsection{RQ1: Overall Performance on QA Tasks}
\label{sec:rq1}

\textbf{Software documentation provides essential value for repository comprehension.}
As illustrated in Tables~\ref{tab:rq1_results}, the ``No Doc'' setting (i.e., addressing tasks without inquiring documentation) achieves average performance of only 48.68 and -3.43 in B-ACC and MCC, 30.34 and 28.49 in F1 and IoU, and 18.40 and 20.74 in $\text{EM}_{1.0}$ and $\text{EM}_{0.8}$, respectively. It highlights the challenging nature of \benchmark, where advanced LLMs cannot effectively answer repository-level questions from prior knowledge without consulting the documentation. All six selected methods consistently outperform the ``No Doc'' setting, with absolute improvements ranging from 5.39\% to 16.22\%, 13.03\% to 32.77\%, 17.45\% to 37.95\%, 16.61\% to 36.41\%, 5.97\% to 9.69\%, and 6.39\% to 10.54\% across these six metrics. This directly confirms that documentation offers indispensable value for understanding and locating repository functionalities. 

\textbf{Despite recent advancements, current documentation generation methods still exhibit great limitations.} Our experiments reveal that even top-performance methods struggle to achieve high scores. In functionality detection, the best performance method achieves an average MCC of only 29.35. While performance on functionality localization is relatively higher, with the leading method achieving an average IoU of 64.90, this still implies a notable localization deviation. This limitation is most pronounced in functionality completion, where the leading method scores merely 28.09 and 31.28 on $\text{EM}_{1.0}$ and $\text{EM}_{0.8}$, on average. These results collectively indicate that current automatically generated documentation demonstrates limited navigation ability and struggles to provide comprehensive details. Thus, there is still a gap between current automated methods and practical developer usage.
   
\begin{myfindbox}
    \textbf{Finding 1:} Although documentation aids repository comprehension, the limited performance of current automated methods constrains their practical value for developers.
\end{myfindbox}

\begin{figure*}[t]
	\centering
	\includegraphics[width=\textwidth]{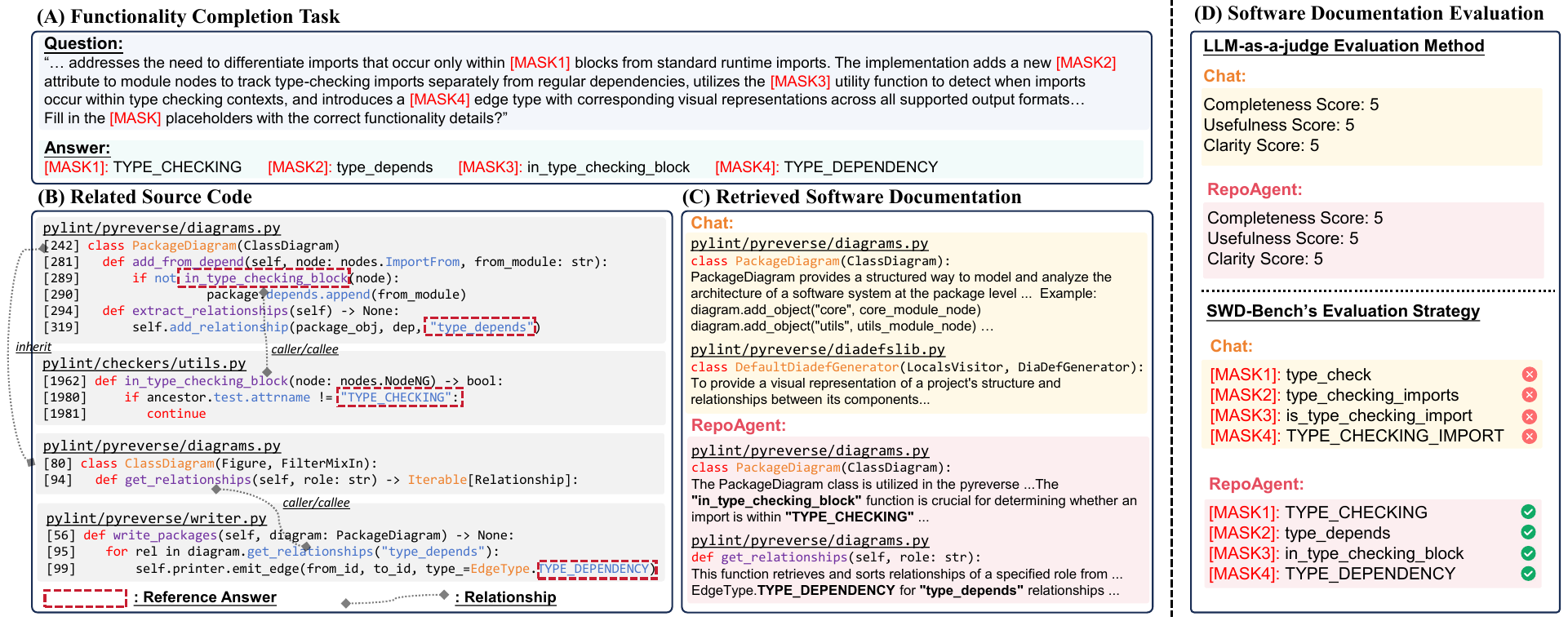}
    \vspace{-2.7em}
    \caption{A case study on the functionality detail task (Entry-ID: ``\texttt{pylint-dev/pylint/8824}''). (A) The task question and answer. (B) The related source code to implement the described functionality. (C) The retrieved documentation from Chat and RepoAgent. (D) A comparison of results between the current evaluation method and our evaluation strategy.}
\label{fig:case_study}
\vspace{-1em}
\end{figure*}

\textbf{Fine-grained documentation generation methods achieve superior performance.} We categorize the four repository-level documentation generation methods by their generated documentation granularity: fine-grained (RepoAgent and DocAgent, which focus on code snippets), intermediate-grained (AutoDoc, which operates at the file level), and coarse-grained (DeepWiki, which produces module-level summaries). Overall, fine-grained methods achieve better performance. Specifically, the average performance of RepoAgent and DocAgent relatively improves upon the AutoDoc by 9.45\% and 65.04\% in B-ACC and MCC, 9.59\% and 9.87\% in F1 and IoU, 9.98\% and 8.08\% in $\text{EM}_{1.0}$ and $\text{EM}_{0.8}$. Furthermore, AutoDoc demonstrates a relative improvement over the DeepWiki of 4.04\%, 49.58\%, 26.91\%, 27.19\%, 2.17\%, and 3.34\% on these six metrics. This suggests that fine-grained documentation provides more comprehensive and concrete information, including parameter usage and code examples, which greatly aid in understanding functionality.

\textbf{Integrating global semantic context is crucial for generating high-quality documentation.} Chat, DocAgent, and RepoAgent all generate fine-grained documentation, but they adopt different strategies for integrating global semantic context, resulting in notable performance differences. RepoAgent populates the prompt with comprehensive context using a repository-wide DAG, DocAgent relies on a ``searcher agent'' to retrieve relevant context, and Chat serves as a baseline comparison without context integration. As shown in Tables~\ref{tab:rq1_results}, DocAgent relatively improves upon Chat by 2.86\%, 30.34\%, 3.95\%, 4.02\%, 4.15\%, and 3.56\%, demonstrating that integrating global context directly enhances documentation quality. 
RepoAgent relatively improves upon DocAgent by 11.45\%, 62.27\%, 5.61\%, 6.10\%, 5.32\%, and 6.67\% across the six metrics. This indicates that directly integrating global context into the prompt is a more robust strategy than relying on a sub-agent for retrieving, which may introduce information incompleteness. Besides, H-Written (human-written documentation artifacts) outperform Chat and is even competitive with DocAgent. For instance, its performance on the functionality localization task under the Gemini-2.5-pro model achieves relative improvements of 2.49\% in F1 and 3.43\% in IoU over DocAgent. This result reveals the authentic process of developers' documentation generation, where developers naturally incorporate repository-level context when writing documentation.

\begin{myfindbox}
    \textbf{Finding 2:} Fine-grained methods that utilize comprehensive context deliver stronger performance. Human-written documentation remains competitive, likely because developers naturally incorporate repository-level context.
\end{myfindbox}

\textbf{Extensive documentation-based inquiry enhances deeper repository comprehension.}
To simulate different levels of developer engagement with documentation, we configure three retrieval settings: brief overview (Top 1024 tokens), standard review (Top 2048 tokens), and in-depth inspection (Top 4096 tokens). Our results demonstrate that accessing more documentation information consistently leads to better outcomes. Specifically, transitioning from a brief overview to a standard review, the average performance of six methods relatively improves by 2.71\%, 21.31\%, 4.31\%, 4.10\%, 2.22\%, and 1.65\% on the six metrics. Further expanding to an in-depth inspection yields additional relative improvements of 2.02\%, 11.24\%, 4.03\%, 3.71\%, 1.86\%, and 1.81\%. This trend mirrors the documentation-driven development, where deeper documentation reading results in more accurate repository understanding.

\textbf{\benchmark provides stable evaluation across different foundational models.} Our experiments demonstrate that while the choice of foundational model influences absolute scores, the relative ranking of documentation generation methods remains consistent. Specifically, Gemini-2.5-pro outperforms GPT-4.1 by 5.04\% on B-ACC, 2.57\% on F1, and 2.95\% on $\text{EM}_{1.0}$ on average, likely due to its advanced comprehension capabilities. However, across all methods, RepoAgent consistently ranks first, followed by DocAgent, AutoDoc, and DeepWiki. This indicates that \benchmark can reliably evaluate the documentation under different models. 

\begin{myfindbox}
    \textbf{Finding 3:} Extensive and in-depth documentation-based inquiry enhances repository comprehension. Besides, our evaluation strategy remains stable under different models.
\end{myfindbox}

\subsection{RQ2: Effectiveness of \benchmark's Evaluation Strategy}
We present a case study to demonstrate the advantages of our evaluation strategy over traditional LLM-as-a-judge evaluation. As illustrated in Figure~\ref{fig:case_study}, we select a functionality completion task from \benchmark (Entry-ID: \texttt{pylint-dev/pylint/8824}). The task question (with masked details) and reference answer are presented in Figure~\ref{fig:case_study} (A). Completing this task requires extracting precise and fine-grained information cross multi files. Figure~\ref{fig:case_study} (B) shows the source code for implementing the described functionality, with technical details highlighted in red boxes representing the ground truth for the reference answers. This functionality involves complex cross-file interactions: for example, the ``\texttt{in\_type\_checking\_block}'' function in ``\texttt{utils.py}'' is called by the ``\texttt{add\_from\_depend}'' method in ``\texttt{diagrams.py}'' (line 289), and the ``\texttt{get\_relationships}'' method in ``\texttt{diagrams.py}'' (line 94) is invoked by ``\texttt{writer.py}'' (line 95). Correctly answering the question requires a holistic understanding of these interactions.

Figure~\ref{fig:case_study} (C) displays the retrieved documentation from Chat and RepoAgent. RepoAgent's documentation contains the necessary details (highlighted in bold) to answer the question, due to its integration of global semantic context during generation. For instance, its documentation for the ``\texttt{PackageDiagram}'' class introduces the concept of ``\texttt{TYPE\_CHECKING}'' from the ``\texttt{in\_type\_checking\_block}'' function, and its explanation of ``\texttt{get\_relationships}'' covers the ``\texttt{TYPE\_DEPENDENCY}'' edge type from the ``\texttt{write\_packages}'' function. This context-aware approach enables developers to better understand the implementation and interaction of specific functionalities.
In contrast, Chat's documentation, which lacks repository-level context, produces only generic descriptions with limited guidance. Figure~\ref{fig:case_study} (D) illustrates the results of two different evaluation strategies. The LLM-as-a-judge method assesses the documentation on dimensions like ``Completeness'' and ``Usefulness''. Since the LLM lacks prior knowledge of the repository, it can only evaluate surface-level quality. As a result, it awards both documentation a perfect score of 5, failing to distinguish their practical value. In contrast, \benchmark's evaluation strategy, based on repository-level QA tasks, reveals clear differences: RepoAgent correctly fills the four masked placeholders, while Chat fails on all of them. This case study demonstrates that our evaluation strategy can assess the practical guidance of software documentation for development.

\begin{myfindbox}
    \textbf{Finding 4:} Compared to current evaluation methods, our evaluation strategy based on functionality-driven QA tasks can provide an accurate assessment of documentation quality. 
\end{myfindbox}

\subsection{RQ3: Impact of Documentation Quality on Issue Solving}

\begin{figure}[t]
    \centering
    \includegraphics[width=.45\textwidth]{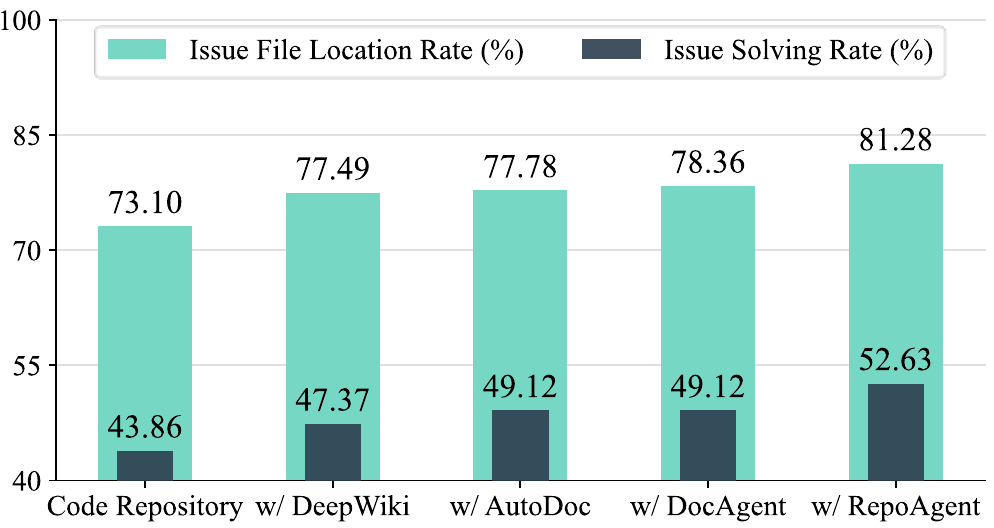}
    \vspace{-1.2em}
    \caption{Performance of SWE-Agent on issue solving when provided with different software documentation.}
    \label{fig:discussion}
\end{figure}

In this RQ, we investigate the impact of documentation quality on issue solving. Based on the selected version in Section~\ref{sec:repo_selection}, we collect 57 corresponding instances from the SWE-Bench Verified~\cite{swebench} and adopt SWE-Agent~\cite{swe-agent} as a representative issue-solving method. For each instance, we provide SWE-Agent with retrieved software documentation (Top 4096 tokens) based on the issue description. The results are shown in Figure~\ref{fig:discussion}. 

\textbf{Software documentation can help improve issue-solving performance.} The baseline issue-solving rate of SWE-Agent (retrieving from the code repository) is 43.86\%. When documentation is provided, the relative issue-solving improvement ranges from 8.00\% to 20.00\%, with the issue file location rates improving by 6.01\% to 11.19\%. These enhancements can be attributed to the global information and complementary context provided in the documentation, which helps SWE-Agent locate and address issues. 

\textbf{Higher-quality documentation provides greater benefits in issue-solving.} The performance ranking of four repository-level documentation generation methods observed in RQ1—with RepoAgent ranks highest, followed by DocAgent, AutoDoc, and DeepWiki—is similarly reflected in the issue-solving results. Specifically, RepoAgent achieves the highest issue-solving rate at 52.63\%, followed by DocAgent and AutoDoc, both at 49.12\%, and DeepWiki at 47.37\%. This consistency demonstrates that our evaluation strategy is effective for evaluating documentation, as higher-quality documentation aids in solving real-world issues.

\begin{myfindbox}
\textbf{Finding 5:} Higher-quality software documentation is more conducive to issue solving, highlighting its practical value in supporting documentation-driven development.
\end{myfindbox}

%% file: Sections/6_Discussion.tex
\input{Tables/RQ2}

\subsection{Complementary Value of Source Code}

We further evaluate the complementary value of source code to software documentation on repository comprehension. We design two settings: (1) inquiring only the documentation generated by the best-performance method (RepoAgent),  (2) inquiring both documentation and source code, with code segmented and embedded by syntax structure and context windows evenly allocated. 

As shown in Table~\ref{tab:rq2_results}, the combined approach consistently outperforms inquiring documentation alone. Specifically, it achieves average relative improvements of 10.39\% and 40.35\% for functionality detection, 12.43\% and 12.88\% for functionality localization, and 2.52\% and 3.62\% for functionality completion. These results highlight that source code, by providing direct implementation details, is essential for enhancing repository-level comprehension. Besides, we find that this complementary value is task-dependent. The synergy between documentation and code is most pronounced in functionality detection and localization, with maximum absolute improvements of 8.91\% and 15.50\% for detection, and up to 9.45\% and 9.46\% for localization. This effectiveness stems from the code's ability to supply precise information. However, this synergy is less evident for functionality completion. For instance, under the GPT-4.1 model and Top 1024 tokens context, performance drops slightly by 0.18\% and 0.26\% in $\text{EM}_{1.0}$ and $\text{EM}_{0.8}$, respectively. This indicates that, within a brief overview, broader global information from documentation is more valuable for accurate completion.


\subsection{Implication of Findings}

\subsubsection{Implications for Developers}
Developers should regard documentation as a fundamental knowledge source within their development process and utilize automated tools to enhance documentation generation efficiency. It is advisable to prioritize tools that produce fine-grained and context-rich documentation, and to supplement documentation reading with source code for a deeper repository understanding. An effective strategy is to first consult the documentation for a high-level overview and identification of relevant files, followed by detailed code inspection. This strategy also proves effective during issue resolution, helping developers locate and address issues.

\subsubsection{Implications for Researchers}
Current automated documentation generation methods exhibit notable limitations, particularly in providing intent-oriented details that require a global understanding of functionality. Hence, efforts should be directed toward improving documentation’s ability to provide comprehensive implementation details. Besides, integrating global semantic context proves effective, suggesting that exploring diverse strategies for semantic fusion is a promising direction. Further research should also investigate the broader value of documentation across various development scenarios, such as code review and refactoring, to fully uncover its impact throughout the software development process.

\subsection{Threats and Limitations}
One threat is that \benchmark is limited to 12 popular open-source repositories, which may affect the generalizability of our findings. However, our data construction pipeline is extensible, and we intend to incorporate more repositories in the future. Another threat arises from the inherent randomness of LLMs. Since we use LLMs to answer QA tasks, results may vary across trials. To mitigate this, we conduct multiple runs and report the average results.


%% file: Tables/RQ2.tex

\begin{table*}[t]
\centering
\caption{Performance comparison on functionality-driven tasks between the standalone RepoAgent and RepoAgent augmented with source code (RepoAgent + Code).
\textbf{Top XX} indicates the token size of the retrieved documentation and source code context, which is evenly allocated to each source. The largest value in each column is marked in \textbf{bold} and \underline{underlined}.}
\vspace{-1em}

\label{tab:rq2_results}
\setlength{\tabcolsep}{5.5pt}

\begin{tabular}{l|cc|cc|cc|cc|cc|cc}
\toprule
\multirow{2}{*}{\textbf{Model}} & \multicolumn{6}{c|}{\textbf{GPT-4.1}} & \multicolumn{6}{c}{\textbf{Gemini-2.5-pro}} \\
& \multicolumn{2}{c}{\textbf{Top 1024}} & \multicolumn{2}{c}{\textbf{Top 2048}} & \multicolumn{2}{c|}{\textbf{Top 4096}} & \multicolumn{2}{c}{\textbf{Top 1024}} & \multicolumn{2}{c}{\textbf{Top 2048}} & \multicolumn{2}{c}{\textbf{Top 4096}} \\ \midrule
\rowcolor{mygreen} \multicolumn{13}{c}{\textbf{Functionality Detection}} \\ \midrule
\rowcolor{gray!20} \textbf{Metric (\%)} & B-ACC & MCC & B-ACC & MCC & B-ACC & MCC & B-ACC & MCC & B-ACC & MCC & B-ACC & MCC \\ \midrule
RepoAgent  & 62.19 & 25.92 & 62.97 & 26.74 & 63.75 & 27.77 & 63.28 & 25.04 & 67.66 & 33.38 & 69.53 & 37.23 \\
RepoAgent+Code   & \underline{\textbf{67.81}} & \underline{\textbf{34.80}} & \underline{\textbf{70.47}} & \underline{\textbf{39.33}} & \underline{\textbf{72.66}} & \underline{\textbf{43.27}} & \underline{\textbf{68.13}} & \underline{\textbf{34.32}} & \underline{\textbf{74.53}} & \underline{\textbf{43.52}} & \underline{\textbf{76.25}} & \underline{\textbf{51.89}} \\ \midrule

\rowcolor{myblue} \multicolumn{13}{c}{\textbf{Functionality Localization}} \\ \midrule
\rowcolor{gray!20} \textbf{Metric (\%)} & F1 & IoU & F1 & IoU & F1 & IoU & F1 & IoU & F1 & IoU & F1 & IoU \\ \midrule
RepoAgent & 63.64 & 60.21 & 67.43 & 64.04 & 70.19 & 66.28 & 67.48 & 64.61 & 69.83 & 66.71 & 71.11 & 67.52 \\
RepoAgent+Code & \underline{\textbf{73.09}} & \underline{\textbf{69.62}} & \underline{\textbf{76.65}} & \underline{\textbf{72.94}} & \underline{\textbf{78.22}} & \underline{\textbf{74.58}} & \underline{\textbf{74.49}} & \underline{\textbf{71.03}} & \underline{\textbf{77.75}} & \underline{\textbf{74.39}} & \underline{\textbf{80.39}} & \underline{\textbf{76.98}}\\ \midrule

\rowcolor{mypink} \multicolumn{13}{c}{\textbf{Functionality Completion}} \\ \midrule
\rowcolor{gray!20} \textbf{Metric (\%)} & $\text{EM}_{1.0}$ & $\text{EM}_{0.8}$ & $\text{EM}_{1.0}$ & $\text{EM}_{0.8}$ & $\text{EM}_{1.0}$ & $\text{EM}_{0.8}$ & $\text{EM}_{1.0}$ & $\text{EM}_{0.8}$ & $\text{EM}_{1.0}$ & $\text{EM}_{0.8}$ & $\text{EM}_{1.0}$ & $\text{EM}_{0.8}$ \\
\midrule
RepoAgent & \underline{\textbf{26.40}} & \underline{\textbf{29.37}} & 26.06 & 28.53 & 26.87 & 29.72 & 29.40 & \underline{\textbf{33.20}} & 29.76 & 32.95 & 30.05 & 33.88 \\
RepoAgent+Code  & 26.22 & 29.11 & \underline{\textbf{26.61}} & \underline{\textbf{30.62}} & \underline{\textbf{28.41}} & \underline{\textbf{31.88}} & \underline{\textbf{29.50}} & 33.13 & \underline{\textbf{30.73}} & \underline{\textbf{34.36}} & \underline{\textbf{31.31}} & \underline{\textbf{35.34}}\\
\bottomrule
\end{tabular}
\end{table*}

%% file: Sections/7_Related_Work.tex
\subsection{Automatic Software Documentation Generation}
Automatic software documentation generation methods can be categorized into three types. Template-based methods~\cite{template2, template3, template4} parse specific information from source code and then populate it into predefined templates. For instance, Hill et al.~\cite{template1} generate annotations by analyzing the identifiers of Java methods. Information retrieval-based methods retrieve suitable descriptions from a vast documentation corpus~\cite{info-retrieve1, info-retrieve4, info-retrieve5}, including bug tracking systems~\cite{info-retrieve2} and developer forums like Stack Overflow~\cite{info-retrieve3}. Deep learning-based methods represent a major focus of current research~\cite{deep-learning1, deep-learning2, deep-learning5}. DocAgent~\cite{docagent} designs a multi-agent framework to generate high-quality documentation. RepoAgent~\cite{repoagent} utilizes global context to infer code functionality and semantics. 

\subsection{Software Documentation Evaluation}
Existing evaluation methods for software documentation can be classified into three categories. Human-based methods~\cite{human1, human2, human3} invite experts to provide detailed assessment, which is labor-intensive. Metrics-based methods~\cite{bleu, meteor, cider, rouge} borrow metrics from Natural Language Processing (NLP), focusing on quantifying the textual similarity between the generated documentation and the references. However, these methods typically rely on high-quality reference documentation, which is quite challenging to construct. Nowadays, LLM-as-a-judge methods have gained traction~\cite{hierarchical, docagent} by leveraging the contextual understanding and instruction-following capabilities of LLMs. By providing LLMs with evaluation criteria, they can conduct assessments across different dimensions. 

%% file: Sections/8_Conclusion.tex
In this paper, we introduce \benchmark, a novel benchmark for evaluating repository-level software documentation generation. We conduct in-depth experiments on this benchmark with several documentation generation methods, conclude our findings, and provide insights for developers and researchers.
To conclude, \benchmark provides a reliable foundation for advancing higher-quality and practical automated documentation generation methods.

